\documentclass{article}

\usepackage{PRIMEarxiv}
\usepackage[utf8]{inputenc}
\usepackage[T1]{fontenc}
\usepackage{url}
\usepackage{booktabs}
\usepackage{amsmath}
\usepackage{amsfonts}
\usepackage{amssymb}
\usepackage{microtype}
\usepackage{fancyhdr}
\usepackage{graphicx}
\usepackage{float}
\usepackage{array}
\usepackage{tabularx}
\usepackage{makecell}
\usepackage{algorithm}
\usepackage{algpseudocode}
\usepackage[nocompress]{cite}
\makeatletter
\renewcommand{\citepunct}{,\penalty\@m}
\makeatother
\usepackage{hyperref}
\hypersetup{
  hidelinks,
  pdftitle={SAEFUZZ: SMART CONTRACT VULNERABILITY DETECTION THROUGH STATICALLY GUIDED EVOLUTIONARY FUZZING},
  pdfauthor={Shiting Yu, Rundong Wei, Xiaoqi Li},
  pdfkeywords={smart contract, fuzz testing, static analysis, vulnerability detection, evolutionary testing}
}

\title{SAEFUZZ: SMART CONTRACT VULNERABILITY DETECTION THROUGH STATICALLY GUIDED EVOLUTIONARY FUZZING}

\author{
  \begin{tabular}{ccc}
    \textbf{Shiting Yu}\thanks{These authors contributed equally to this work.} & \textbf{Rundong Wei}\textsuperscript{*} & \textbf{Xiaoqi Li} \\
    \textnormal{Hainan University} & \textnormal{Hainan University} & \textnormal{Hainan University} \\
    \textnormal{Haikou, China} & \textnormal{Haikou, China} & \textnormal{Haikou, China} \\
    \texttt{3299084399@qq.com} & \texttt{weirundong2026@hainanu.edu.cn} & \texttt{csxqli@ieee.org}
  \end{tabular}
}

\begin{document}
\maketitle

\begin{abstract}
The effectiveness of smart contract fuzzing depends strongly on whether generated transactions reach deep, state-dependent execution paths. Existing fuzzers often generate highly random call sequences, wasting executions on semantically invalid or low-value states and leaving vulnerabilities that require specific invocation orders unexplored. We present a lightweight method for generating fuzz test cases under bytecode-level static guidance. We construct an Ethereum virtual machine control-flow graph, extract paths containing vulnerability-relevant instructions, recover function selectors, and order externally callable functions according to storage read--write dependencies. A coverage-guided evolutionary strategy then generates, evaluates, recombines, and mutates executable seeds. Five dedicated runtime oracles target reentrancy, integer overflow or underflow, block-state dependence, unsafe delegate calls, and frozen Ether. The evaluation uses deployed Ethereum contracts, including labelled vulnerable contracts. SAEFUZZ detects most labelled vulnerable contracts, yielding 98.50\% accuracy, 90.00\% precision, and 81.82\% recall. It also achieves 84.07\% mean instruction coverage, with valid test cases accounting for 93.48\% of generated cases. Ablation results indicate that static guidance, directed seed generation, and vulnerability-specific oracles each contribute to the final performance.
\end{abstract}

\keywords{Smart Contract \and Fuzz Testing \and Static Analysis \and Vulnerability Detection}

\section{Introduction}

Smart contracts provide the executable logic for digital asset management and decentralised applications. Once deployed, their code and persistent state are difficult to modify, and their functions may control valuable assets. Vulnerabilities that survive deployment can therefore cause irreversible state corruption or financial loss. Manual review remains important, but the growing size and complexity of contract code motivate automated analysis before deployment. Fuzz testing is attractive in this setting because it validates contract behaviour through concrete execution. A fuzzer repeatedly submits generated transactions, observes execution traces and state changes, and searches for behaviours that satisfy a vulnerability oracle. Compared with purely static pattern matching, execution can reduce alarms that are infeasible at runtime. However, a smart contract is stateful: the effect of a call depends on earlier transactions, storage values, caller identities, transferred Ether, and environmental data. Random single calls or unordered sequences often remain in shallow states and consume much of the testing budget on invalid inputs.

We call this method SAEFUZZ, a statically guided evolutionary fuzzing method for smart contract vulnerability detection. It combines bytecode-level static analysis with evolutionary fuzzing. Static analysis identifies vulnerability-relevant control-flow paths and extracts externally callable function sequences whose ordering is consistent with storage dependencies. These sequences provide structured starting points for fuzzing. Evolutionary operators then optimise the corresponding test cases using instruction and branch coverage, while dedicated test oracles inspect runtime traces for five high-risk vulnerability classes. We deliberately avoid symbolic execution because constraint solving and path exploration can be expensive on contracts with large state spaces.

The main contributions are as follows:
\begin{enumerate}
  \renewcommand{\labelenumi}{(\arabic{enumi})}
  \item We present a transaction sequence extraction procedure that builds an EVM control-flow graph, retains paths containing security-relevant opcodes, recovers function selectors, and prioritises externally callable state-changing sequences.
  \item We develop an evolutionary test case generator that constructs type-valid seeds from contract ABIs and optimises them using branch and instruction coverage through selection, crossover, and mutation.
  \item We define runtime oracles for reentrancy, arithmetic overflow or underflow, block-state dependence, unsafe delegate calls, and frozen Ether, and integrate them into an end-to-end detection workflow.
  \item We evaluate SAEFUZZ on deployed Ethereum contracts and analyse detection quality, code coverage, the ratio of valid test cases, and three module ablations.
\end{enumerate}

\section{Related Work and Background}

\subsection{Smart Contract Fuzzing}

ContractFuzzer introduced ABI-guided input generation and vulnerability-specific test oracles for Ethereum contracts \cite{Jiang2018ContractFuzzer}. Later systems target different parts of the fuzzing loop. sFuzz applies an adaptive strategy to improve branch exploration, while ILF combines symbolic execution with imitation learning to learn a fuzzing policy. ItyFuzz uses dataflow-guided snapshots to accelerate the exploration of stateful exploits. Fuzzing methods that account for invocation order revisit important branches to improve the exploration of deep states \cite{Liu2023InvocationOrdering}, while Harvey demonstrated demand-driven greybox fuzzing for multi-transaction smart contract testing \cite{Wustholz2020Harvey}. MuFuzz further combines sequence-aware mutation with seed mask guidance \cite{Qian2024MuFuzz}, and EF/CF compiles EVM bytecode to native code and applies structure-aware mutation for high-throughput exploit generation \cite{Rodler2023EFCF}. These systems collectively indicate that invocation order, state feedback, and mutation guidance are central to effective contract fuzzing. Other program analysis approaches complement dynamic testing. S-gram uses semantic information for Ethereum security auditing \cite{Liu2018Sgram}, HoRStify provides sound static security analysis \cite{Holler2023HoRStify}, and SmartInv learns contract invariants from multimodal representations \cite{Wang2024SmartInv}. Large language models have also been used as test generators or mutators. TitanFuzz combines generative and infilling models with evolutionary fuzzing for deep learning libraries \cite{Deng2023TitanFuzz}, and structured prompting has been studied for seeds for parser fuzzing \cite{Ackerman2023LLMFuzzing}. Cross-contract exploit analysis and interaction-aware bytecode analysis further suggest that important behaviours can be missed when contracts are analysed in isolation \cite{Li2025DeFiExploits,Li2025InteractionAware}. Although semantic models can produce context-sensitive inputs, they add inference cost and may require manual validation. We instead examine whether a lightweight combination of static dependency analysis, evolutionary testing, and explicit runtime oracles can improve smart contract fuzzing.\par
Recent learning-based detectors combine source code, bytecode, graph, and language model representations. Multimodal detection, LLM-enhanced analysis, multi-agent auditing, and reinforcement-tuned explanations have been explored in complementary systems \cite{Xiong2026Multimodal,Ding2025SmartGuard,Wei2025MultiAgent,Yu2025SmartLLMADPO}, while other approaches study multimodal agents, deep learning representations, universal big data pipelines, and graph neural networks for contract-level classification \cite{Jie2025Agent4Vul,Xiong2025NDLSC,Lian2025MultiModal,Xu2025GNN}. Related work also investigates multiscale fusion, graph-based contract sensing, deep feature fusion, and combinations of language models with conventional machine learning \cite{Wang2025TMFNet,Pang2025ContractSensing,Chu2025DeepFusion,Hossain2025LLMML}. Security-oriented repair, federated learning, automated analysis frameworks, and knowledge migration address robustness, privacy, or transferability beyond a single detector \cite{Karanjai2025SecureLanguages,Jia2025Federated,Mi2024Automated,Wang2024KnowledgeMigration}. Evolution-aware analysis of proxy-based upgradeable contracts and non-EVM security studies provide additional evidence that contract evolution and execution context matter for vulnerability analysis \cite{Li2025USCSA,Wu2025Solana}. PSR$^{2}$ detects atomicity violations through phase-based semantic reasoning and contract refinement, whereas SCPatcher uses retrieval-augmented generation and a knowledge graph for automated repair \cite{Li2026PSR,li2026scpatcher}. These approaches broaden semantic coverage, but their requirements for training data, model inference, or large-scale preprocessing differ from the lightweight bytecode-guided setting considered here.
Several recent studies frame smart contract security as a lifecycle problem involving detection, evaluation, benchmarking, planning, code transformation, and explanation \cite{Abdelaziz2026BeyondDetection,Wang2026EVMbench,Abdelaziz2026AnalyzerShortfall,Ince2025GenerativeLLM}. Plan-and-execute auditing, transformation-aware analysis, post-training language models, and LLM-based frameworks further broaden the design space \cite{Wei2025PlanExecute,Manh2024CodeTransformation,Yu2024SmartLlama,Zaazaa2024SmartLLMSentry}. Multimodal reasoning for atomicity violations and systematic studies of exploitable patterns offer complementary views of contract-level security evidence \cite{Li2025AtomGraph,Ding2025ExploitablePatterns}. SAEFUZZ addresses a different point in this design space. It uses static analysis to reduce the transaction search space before concrete execution and does not require model inference in the testing loop.
Related studies include machine learning classifiers, heterogeneous graph attention networks, LLM-oriented security surveys, and dedicated reentrancy detectors \cite{Wang2024MachineLearning,Luo2024SCVHUNTER,Boi2024LLMRole,Wang2024Reentrancy}. Survey and evaluation studies compare tools, threat models, datasets, and generalisation limits across smart contract vulnerability detection \cite{Khan2024Survey,Kiani2024Review,Adamantis2025Evaluation,Jiao2024EthereumSecurity}. At the broader systems level, analyses of denial-of-service threats and blockchain hardening clarify security concerns beyond a single EVM detector \cite{Zhang2025DoS,Yang2025BlockchainSecurity}. These comparisons reinforce the need to report not only detection counts but also test validity, coverage, false alarms, and the assumptions underlying each benchmark.\par
Other representation learning and program analysis studies consider security analysis taxonomies, contrastive learning, version-aware tracing, and dual-view contract representations \cite{Zhu2024Survey,Chen2024Contrastive,Mbodji2024ContractTrace,Yao2024DualView}. Deep learning approaches have also been applied to vulnerability classification and transfer learning, including Ethereum-specific detection pipelines \cite{Patel2024DeepLearning,Tang2023DeepLearning}. Ethereum implementation security and cryptography-oriented blockchain reviews provide broader context for the protocol assumptions underlying smart contract analysis \cite{Gao2025EthereumCrypto,Zhou2025CryptoReview}. The SAEFUZZ workflow complements these classifiers because its primary output is an executable, vulnerability-oriented transaction sequence together with trace evidence.
Within smart contract security, cross-contract exploit detection, interaction-aware bytecode analysis, evolution-aware analysis of proxy-based upgradeable contracts, and non-EVM security studies provide adjacent evidence about the limits of isolated syntax \cite{Li2025DeFiExploits,Li2025InteractionAware,Li2025USCSA,Wu2025Solana}. Ethereum-focused detection and deep transfer learning pipelines are also relevant comparison points for vulnerability classification \cite{Yang2023Ethereum,Sendner2023DeepTransfer}. A systematic survey of DeFi composability further connects code correctness with protocol robustness \cite{li2026systematic}. Phase-based semantic reasoning for atomicity violation detection, knowledge-graph-assisted repair, multimodal atomicity reasoning, and systematic studies of exploitable patterns further motivate combining structural evidence with semantic context \cite{Li2026PSR,li2026scpatcher,Li2025AtomGraph,Ding2025ExploitablePatterns}.
Broader blockchain security studies examine denial-of-service threats, system hardening, Ethereum implementation security, and cryptographic foundations \cite{Zhang2025DoS,Yang2025BlockchainSecurity,Gao2025EthereumCrypto,Zhou2025CryptoReview}. These works define scope boundaries for the present evaluation: SAEFUZZ targets five EVM-level vulnerability classes and does not claim coverage of every protocol, cryptographic, or availability threat. Adjacent security contexts include autonomous agents that invoke tools and library misuse detection using iterative feedback and static verification \cite{Li2026OpenClaw,wang2026libscan}. Exploitable pattern studies and knowledge-graph-assisted access-control detection provide further examples of the need for transaction-level evidence and structured security reasoning \cite{Ding2025ExploitablePatterns,li2026ckg}. These studies are outside the evidence base for the reported fuzzing results, but they motivate stronger provenance, structured outputs, and explicit boundaries between detection evidence and broader security reasoning in future auditing systems.

\subsection{EVM Control and Data Flow}

Solidity contracts compile to EVM bytecode. The EVM is a stack machine whose execution state includes a program counter, operand stack, transient memory, persistent storage, and remaining gas. A control-flow graph (CFG) represents bytecode execution as basic blocks connected by directed edges. A basic block begins at an entry point and ends at a control-transfer or terminating instruction such as \texttt{JUMP}, \texttt{JUMPI}, \texttt{STOP}, \texttt{REVERT}, or \texttt{SELFDESTRUCT}. Data-flow analysis then tracks how stack values and storage locations are produced, propagated, read, and updated. For stateful fuzzing, storage dependencies provide a useful approximation of semantics between functions. A function that executes \texttt{SSTORE} can establish a state required by a later function that executes \texttt{SLOAD}. Placing a call that writes state before a call that reads state does not guarantee semantic validity, but it removes many sequences that are unlikely to represent meaningful state transitions. SAEFUZZ uses this dependency as a lightweight ordering heuristic and then validates generated inputs through concrete execution.

\subsection{Target Vulnerabilities}

We focus on the five vulnerability classes in Table~\ref{tab:vulnerabilities}. We select them because they exhibit observable opcode, state, or asset-flow patterns that can be monitored during fuzzing.

\begin{table}[H]
\centering
\caption{Target vulnerabilities and monitored characteristics.}
\label{tab:vulnerabilities}
\begin{tabularx}{\textwidth}{@{}lXX@{}}
\toprule
Vulnerability & Primary impact & Monitored characteristic \\
\midrule
Reentrancy & Asset theft or inconsistent state & External call followed by a delayed update to the same storage location \\
Integer overflow/underflow & Corrupted balances or numeric state & Unprotected arithmetic result outside the intended range \\
Block-state dependence & Manipulable contract decision & \texttt{TIMESTAMP} or \texttt{NUMBER} affects an Ether transfer \\
Unsafe delegate call & Storage corruption or contract takeover & Attacker-controlled \texttt{DELEGATECALL} target or arguments \\
Frozen Ether & Permanently locked funds & Contract accepts Ether but cannot complete an outward transfer \\
\bottomrule
\end{tabularx}
\end{table}

\section{Transaction Sequence Extraction}

\subsection{Framework Overview}

\begin{figure}[H]
\centering
\includegraphics[width=0.95\linewidth]{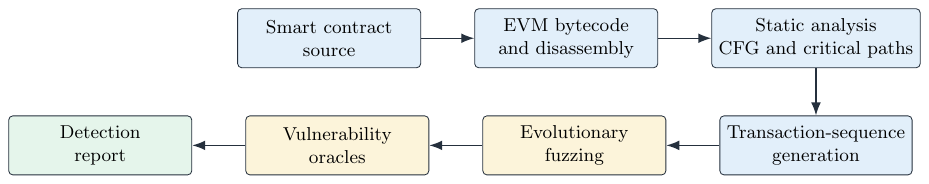}
\caption{Statically guided fuzzing workflow.}
\label{fig:framework}
\end{figure}

The SAEFUZZ framework contains three stages, as illustrated in Figure~\ref{fig:framework}. First, the sequence extraction stage compiles and disassembles a contract, builds its CFG, identifies paths containing vulnerability-relevant instructions, and constructs ordered transaction sequences. Second, the seed evolution stage instantiates ABI-valid inputs and iteratively improves them using coverage feedback. Third, the detection stage executes evolved tests and applies vulnerability-specific oracles to the resulting traces and state changes.

\subsection{Critical Path Extraction}

Let $G=(V,E)$ denote the CFG of a contract, where each vertex is a basic block. We define a set $K$ of instructions associated with the target vulnerabilities. Table~\ref{tab:key-opcodes} summarises the principal mapping. Arithmetic and environment opcodes identify candidate operations. Storage and call opcodes provide contextual evidence for sequence construction and runtime validation.

\begin{table}[H]
\centering
\caption{Security-relevant opcode set used for path extraction.}
\label{tab:key-opcodes}
\begin{tabular}{@{}lll@{}}
\toprule
Opcode & Associated risk & Priority \\
\midrule
\texttt{CALL} & Reentrancy or Ether transfer & High \\
\texttt{SSTORE}, \texttt{SLOAD} & State-dependent behaviour & Medium \\
\texttt{ADD}, \texttt{SUB}, \texttt{MUL} & Integer overflow/underflow & High \\
\texttt{TIMESTAMP}, \texttt{NUMBER} & Block-state dependence & Medium \\
\texttt{DELEGATECALL} & Unsafe delegated execution & High \\
\bottomrule
\end{tabular}
\end{table}

Algorithm~\ref{alg:path} marks every block containing an instruction in $K$. Depth-first traversal then retains complete paths that contain at least one marked block. This step reduces the path set passed to later processing. It does not itself classify a vulnerability.

\begin{algorithm}[H]
\caption{Critical path extraction}\label{alg:path}
\begin{algorithmic}[1]
\Require Contract bytecode $B$, critical instruction set $K$
\Ensure Candidate critical paths $P$
\State $G \gets \Call{BuildCFG}{B}$
\State $P \gets \emptyset$
\ForAll{basic blocks $v \in G$}
  \If{$v$ contains an instruction in $K$}
    \State mark $v$ as critical
  \EndIf
\EndFor
\ForAll{marked blocks $v$}
  \ForAll{complete paths $p$ reached by \Call{DFS}{$G,v$}}
    \If{$p$ contains at least one marked block}
      \State $P \gets P \cup \{p\}$
    \EndIf
  \EndFor
\EndFor
\State \Return $P$
\end{algorithmic}
\end{algorithm}

\subsection{Function Recovery and Sequence Construction}

Externally callable Solidity functions are dispatched through four-byte selectors derived from their signatures. The dispatcher commonly compares call data against constants loaded by \texttt{PUSH4}. For each critical path, SAEFUZZ therefore scans for \texttt{PUSH4} operands, maps them to ABI entries, and records recovered functions in path order. Candidate lists are deduplicated and combined into transaction sequences. Storage operations impose a lightweight dependency order. Functions whose paths contain \texttt{SSTORE} are placed before functions whose paths contain \texttt{SLOAD}, because the former may establish persistent state consumed by the latter. A public or external function containing \texttt{SSTORE} contributes two priority points. A function containing \texttt{SLOAD} contributes one. Sequences are sorted in descending order of the accumulated score. The visibility filter removes calls that an external adversary cannot invoke. Read-only functions are deprioritised because they cannot modify persistent state, although they may still be useful as observation calls in a more general sequence model. This simplification concentrates the available fuzzing budget on sequences with a higher estimated probability of changing security-relevant state.

\begin{algorithm}[H]
\caption{Transaction sequence extraction and prioritisation}\label{alg:sequence}
\begin{algorithmic}[1]
\Require Critical paths $P$, contract ABI $A$
\Ensure Prioritised sequence list $S$
\State $S \gets \emptyset$
\ForAll{paths $p \in P$}
  \State $F \gets \emptyset$
  \ForAll{instructions $i$ on $p$}
    \If{$i.opcode=\texttt{PUSH4}$}
      \State append $\Call{ResolveABI}{i.operand,A}$ to $F$
    \EndIf
  \EndFor
  \State $S \gets S \cup \Call{CombineAndDeduplicate}{F}$
\EndFor
\ForAll{sequences $s \in S$}
  \State $s.score \gets 0$
  \ForAll{functions $f \in s$}
    \If{$f$ is public or external and contains \texttt{SSTORE}}
      \State $s.score \gets s.score+2$
    \ElsIf{$f$ is public or external and contains \texttt{SLOAD}}
      \State $s.score \gets s.score+1$
    \EndIf
  \EndFor
\EndFor
\State remove functions that are private, internal, \texttt{view}, \texttt{pure}, or legacy \texttt{constant}
\State \Return sequences sorted by decreasing score
\end{algorithmic}
\end{algorithm}

\section{Test Case Generation and Vulnerability Oracles}

\subsection{Initial Seed Pool}

Each prioritised transaction sequence is instantiated using the function descriptions in the ABI. Fixed-width types such as unsigned integers and Booleans are sampled from their legal domains. Ethereum addresses are encoded as 20-byte values. For dynamic arrays and byte arrays, the generator first samples a length and then populates elements within the corresponding type range. These rules are intended to produce executable initial inputs while retaining enough diversity for later mutation.

\subsection{Fitness Evaluation and Evolution}

\begin{figure}[H]
\centering
\includegraphics[height=0.43\textheight]{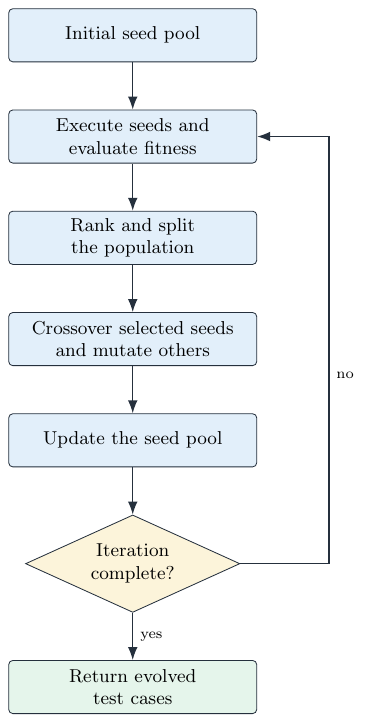}
\caption{Coverage-guided seed evolution workflow.}
\label{fig:evolution}
\end{figure}

Let $E_{br}(s)$ be the number of branch outcomes reached by seed $s$, and let $B_{jump}$ be the total number of branch outcomes tracked for the contract. Branch coverage is
\begin{equation}
C_{br}(s)=\frac{E_{br}(s)}{B_{jump}}.
\end{equation}
Let $E_{pc}(s)$ be the number of executed instruction positions and $P_{code}$ the total number of instruction positions. Instruction coverage is
\begin{equation}
C_{code}(s)=\frac{E_{pc}(s)}{P_{code}}.
\end{equation}
We combine the two terms as
\begin{equation}
F(s)=w_{br}C_{br}(s)+C_{code}(s), \qquad
w_{br}=\frac{B_{jump}}{P_{code}}.
\end{equation}
A larger fitness value therefore favours tests that execute more instructions and reach more branch outcomes. Seeds are ranked by fitness. We pair the top fraction $\alpha$ for crossover and recombine transaction subsequences and parameter tuples from two parents. Offspring that violate the ABI or the inferred storage ordering are rejected. Seeds outside the selected fraction undergo mutation with probability $p_m$. Parameter-level mutation replaces values or flips bits. Transaction-level mutation changes call attributes such as the sender or transferred value. Figure~\ref{fig:evolution} illustrates the seed evolution workflow.

\begin{algorithm}[H]
\caption{Seed selection and evolution}\label{alg:evolution}
\begin{algorithmic}[1]
\Require Seed pool $Q$, selection ratio $\alpha$, mutation probability $p_m$
\Ensure Evolved seed pool $Q'$
\ForAll{seeds $s \in Q$}
  \State $s.fitness \gets F(s)$
\EndFor
\State sort $Q$ by decreasing fitness
\State $Q_{top} \gets Q[0:\lfloor\alpha|Q|\rfloor]$
\State $Q' \gets \emptyset$
\State $(\mathcal{P},s_{odd}) \gets \Call{PairAdjacent}{Q_{top}}$
\ForAll{pairs $(s_1,s_2) \in \mathcal{P}$}
  \State $c \gets \Call{Crossover}{s_1,s_2}$
  \If{$\Call{IsValid}{c}$} $Q' \gets Q'\cup\{c\}$ \EndIf
\EndFor
\If{$s_{odd} \ne \emptyset$}
  \State $Q' \gets Q'\cup\{s_{odd}\}$
\EndIf
\ForAll{$s \in Q\setminus Q_{top}$}
  \If{$\Call{Random}{}<p_m$}
    \State $Q' \gets Q'\cup\{\Call{Mutate}{s}\}$
  \Else
    \State $Q' \gets Q'\cup\{s\}$
  \EndIf
\EndFor
\State \Return $Q'$
\end{algorithmic}
\end{algorithm}

\subsection{Vulnerability Oracles}

The oracles combine opcode evidence with runtime state or asset-flow evidence. Table~\ref{tab:oracles} gives their logical structure.
For reentrancy, the trace monitor first checks for a successful \texttt{CALL} that forwards more than a configured gas threshold and transfers a positive value. It then searches storage events for a read--call--write pattern on the same location. This operationalises the checks-effects-interactions violation targeted by our detector. The detector records the test input, call trace, and storage location when both subconditions hold. For arithmetic vulnerabilities, the monitor observes \texttt{ADD}, \texttt{SUB}, \texttt{MUL}, \texttt{DIV}, and \texttt{MOD}, reconstructs the operation from stack operands, and compares the result with the intended integer range. An alarm is emitted only when execution continues without a successful \texttt{require}, \texttt{revert}, or checked arithmetic safeguard. This oracle is most directly applicable to compiler configurations or unchecked regions in which wraparound remains possible. The block-state oracle requires both environmental dependence and asset impact. A use of \texttt{TIMESTAMP} or \texttt{NUMBER} must influence a branch or calculation that controls whether Ether is transferred or determines its value. The delegate call oracle similarly distinguishes the mere presence of \texttt{DELEGATECALL} from an exploitable use by tracing the target and arguments back to unvalidated transaction input. Finally, the frozen Ether oracle first establishes that a positive-value transaction succeeds and then checks whether any outward transfer path can be executed under the generated call sequences.

\begin{table}[H]
\centering
\caption{Runtime test oracles.}
\label{tab:oracles}
\begin{tabularx}{\textwidth}{@{}lX@{}}
\toprule
Vulnerability & Trigger condition \\
\midrule
Reentrancy & $ExistCALL \land DelayedStoreUpdate$: an external call carrying value with sufficient gas occurs, and a storage location read before the call is written after it \\
Integer overflow/underflow & $ExistArith \land ErrorResult \land NoException$: arithmetic occurs, the observed result is inconsistent with the intended range, and no effective revert or checked arithmetic protection handles it \\
Block-state dependence & $(TimestampOp \lor BlockNumberOp) \land TransferEther$: block metadata influences a positive-value Ether transfer \\
Unsafe delegate call & $ExistDelegateCALL \land AttackerControlled$: delegated execution succeeds and its target or arguments are directly controlled by the attacker without validation \\
Frozen Ether & $CanReceive \land NotTransfer$: the contract accepts transactions carrying value but no valid outward Ether transfer can be completed \\
\bottomrule
\end{tabularx}
\end{table}

\subsection{End-to-End Detection}

For each contract, SAEFUZZ compiles the contract code, disassembles the bytecode, builds the CFG, extracts critical paths and transaction sequences, constructs an initial seed pool, and performs a fixed number of evolutionary iterations. Each evolved seed is executed against the contract, and all five oracles inspect the resulting trace. The report records the contract, triggering seed, execution evidence, and vulnerability class.

\section{Evaluation}

\subsection{Dataset and Baselines}

The evaluation includes 200 open-source contracts deployed on Ethereum. The dataset includes ERC-20, ERC-721, ERC-1155, access-control, security component, decentralised finance, and NFT-related contracts. Eleven contracts are labelled with one of the five target vulnerabilities, and 189 are labelled non-vulnerable. Before testing, the contracts were compiled, normalised, checked for executability, and assigned vulnerability labels. We compare SAEFUZZ with Oyente, Osiris, and sFuzz. Oyente applies symbolic execution and constraint solving. Osiris extends an Oyente-style analysis with arithmetic and taint tracking. sFuzz uses evolutionary greybox fuzzing with branch-distance feedback. All experiments ran in a common Windows environment using Python~3.8. The evaluation does not report processor, memory, timeout, baseline version, or repetition details. We discuss these limitations in Section~\ref{sec:limitations}.

\subsection{Detection Results}

Table~\ref{tab:detection} reports the TP, FP, and FN counts. We compute accuracy from these counts and the 200-contract dataset, and compute precision and recall using their standard definitions. SAEFUZZ detects nine of the eleven labelled vulnerable contracts and produces one false positive among the 189 labelled non-vulnerable contracts. Its recall matches that of sFuzz, while its false-positive count is much lower. This pattern is consistent with SAEFUZZ's design: static analysis directs testing toward state-changing sequences, and the runtime oracles require multiple pieces of execution evidence before reporting a vulnerability. The experiment does not establish statistical significance because only 11 positive contracts are present and repeated runs are not reported. SAEFUZZ also achieves 84.07\% mean instruction coverage, with valid test cases accounting for 93.48\% of generated cases.

\begin{table}[H]
\centering
\caption{Detection results on the evaluation dataset.}
\label{tab:detection}
\begin{tabular}{@{}lrrrrrr@{}}
\toprule
Method & TP & FP & FN & Accuracy (\%) & Precision (\%) & Recall (\%) \\
\midrule
Oyente & 5 & 29 & 6 & 82.50 & 14.71 & 45.45 \\
Osiris & 8 & 37 & 3 & 80.00 & 17.78 & 72.73 \\
sFuzz & 9 & 11 & 2 & 93.50 & 45.00 & 81.82 \\
SAEFUZZ & 9 & 1 & 2 & 98.50 & 90.00 & 81.82 \\
\bottomrule
\end{tabular}
\end{table}

\subsection{Ablation Study}

Three ablations remove static guidance, directed generation, or the vulnerability oracle module. Table~\ref{tab:ablation} summarises the ablation results. Removing static guidance produces the largest increase in false positives and reduces detected vulnerabilities from nine to five. Removing directed generation also reduces detections and accuracy, while removing the dedicated oracles sharply increases false positives. Together, these observations indicate that the three modules address different failure modes: reduction of the search space, exploration of executable states, and alarm validation.

\begin{table}[H]
\centering
\caption{Ablation results.}
\label{tab:ablation}
\begin{tabular}{@{}lrrr@{}}
\toprule
Configuration & Detected vulnerabilities & False positives & Accuracy (\%) \\
\midrule
SAEFUZZ & 9 & 1 & 98.5 \\
Without static analysis & 5 & 31 & 82.0 \\
Without directed generation & 6 & 9 & 85.0 \\
Without vulnerability oracles & 7 & 29 & 83.5 \\
\bottomrule
\end{tabular}
\end{table}

\section{Discussion and Limitations}\label{sec:limitations}

The results suggest that SAEFUZZ can improve the precision of evolutionary smart contract fuzzing without adding symbolic execution to the online testing loop. Storage-aware sequence ordering is especially useful when a vulnerability becomes reachable only after a state-changing call. Dedicated oracles further reduce alarms by combining opcode evidence with storage or asset-flow effects rather than treating a suspicious instruction as sufficient evidence on its own. Several limitations constrain the conclusions. First, the positive subset contains only 11 labelled vulnerable contracts, so a small change in detections materially changes recall. The evaluation does not report label adjudication, dataset identifiers, contract addresses, or a public archive, which limits reproducibility and independent verification. Second, baseline versions, command-line options, time budgets, random seeds, hardware specifications, and statistics from repeated runs are not reported. The comparison should therefore be interpreted as an initial evaluation rather than a definitive benchmark. Third, the sequence model uses opcode-level storage reads and writes as a proxy for semantic dependencies. This heuristic may miss dependencies mediated by mappings, dynamic storage layout, cross-contract calls, events, or application-specific invariants. It can also prioritise state changes unrelated to a target vulnerability. Fourth, the oracles are heuristic. Legitimate proxy patterns may resemble unsafe delegate calls. Liveness-based frozen Ether checks depend on how thoroughly withdrawal paths are explored, and arithmetic monitoring must account for Solidity compiler versions and \texttt{unchecked} blocks. Finally, the evaluation covers only five vulnerability classes and EVM-compatible contracts. Extending the method to access-control faults, denial of service, dependence on transaction order, and non-EVM platforms requires new static features, mutators, and runtime evidence. Future work should release the labelled dataset and implementation, specify deterministic experimental protocols, and report runtime and repeated-run variance. Semantic dependency analysis, taint tracking, constraint solving, or language model assistance could be incorporated selectively to improve understanding of business logic while retaining the current lightweight path as a baseline. Cross-platform evaluation on other EVM-compatible chains and non-EVM environments would also clarify which components generalise beyond Ethereum.

\section{Conclusion}

We present SAEFUZZ, a statically guided method for generating fuzz test cases for smart contract vulnerability detection. SAEFUZZ extracts security-relevant bytecode paths, constructs and prioritises transaction sequences using external visibility and storage dependencies, evolves ABI-valid seeds with coverage feedback, and applies runtime oracles for five high-risk vulnerability classes. On the dataset, it detects most labelled vulnerable contracts with one false positive, yielding 98.50\% accuracy, 90.00\% precision, and 81.82\% recall. SAEFUZZ also achieves 84.07\% mean instruction coverage, with valid test cases accounting for 93.48\% of generated cases.

\section*{Acknowledgments}

AI-based tools are used for language polishing during manuscript preparation.

\begingroup
\small
\bibliographystyle{unsrt}
\bibliography{references}
\endgroup

\end{document}